\documentclass[final,1p,times]{elsarticle} 

\usepackage{graphicx}
\usepackage{amssymb} 
\usepackage{amsthm} 
\usepackage{lineno}
\usepackage{bm}

\newcommand{\beq}{\begin{equation}}
\newcommand{\eeq}{\end{equation}}
\renewcommand\r[1]{(\ref{#1})}
\def\({\left(}  \def\){\right)}  \def\[{\left[}  \def\]{\right]}  \def\<{\langle}  \def\>{\rangle}

\journal{Nuclear Physics A} 

\begin{document}

\begin{frontmatter} 

\title{Far-from-equilibrium heavy quark energy loss at strong coupling}


\author{Paul Chesler$^1$}
\author{Mindaugas Lekaveckas$^1$}
\author{Krishna Rajagopal$^{1,2}$}
\address{$^1$Center for Theoretical Physics, Massachusetts Institute of Technology, Cambridge MA 02139}
\address{$^2$Physics Department, Theory Unit, CERN, CH-1211 Gen\`eve 23, Switzerland}

\begin{abstract} 

We study the energy loss of a heavy quark propagating through the matter 
produced in the collision of two sheets of 
energy~\cite{Chesler:2010bi}.   Even though this
matter is initially far-from-equilibrium we find 
that, when written in terms of the energy density, 
the equilibrium expression for heavy quark energy loss 
describes most qualitative
features of our results well. 
At later times, once a plasma described by viscous
hydrodynamics has formed, the equilibrium expression
describes the heavy quark energy loss quantitatively.
In addition to the drag force that makes it  lose energy,
a quark moving through the out-of-equilibrium matter  feels a force
perpendicular to its velocity.

\end{abstract} 

\end{frontmatter} 



The discovery that strongly
coupled quark-gluon plasma (QGP) is produced in ultrarelativistic 
heavy ion collisions has prompted much interest in the
real-time dynamics of strongly coupled non-Abelian plasmas.  
For example, heavy quark energy loss has received substantial attention.  
If one shoots a heavy quark though a 
non-Abelian plasma how much energy does it lose as it propagates?
This question has been answered~\cite{Herzog:2006gh} for 
equilibrium plasma in
strongly coupled $\mathcal N = 4$ supersymmetric Yang-Mills (SYM) theory in the large 
number of colors $N_c$ limit, 
where holography permits a semiclassical description of energy loss in terms of string dynamics in asymptotically AdS$_5$ spacetime \cite{Maldacena:1997re,Karch:2002sh,Herzog:2006gh}.
One challenge (not the only one)
in using these results to glean qualitative insights into 
heavy quark 
energy loss in heavy ion collisions is that a heavy quark produced at time zero
must first propagate through the initially far-from-equilibrium
matter produced in the collision before it later plows through the
expanding, cooling, near-equilibrium strongly coupled QGP. 
In this contribution we sketch our work in progress toward
obtaining guidance for how to meet this challenge.
We want a toy model  in which we can reliably calculate 
how the energy loss rate
of a heavy quark moving through the far-from-equilibrium matter present just after a collision
compares to that in strongly coupled plasma close to equilibrium.

We 
study the energy loss of a heavy quark moving through the debris produced by the collision of
planar sheets of energy in strongly coupled SYM theory analyzed in Ref.~\cite{Chesler:2010bi}.  
The incident sheets of energy move at the speed of light in the $+z$ and $-z$ directions 
and collide at $z = 0$ at time $t= 0$.  They each have a Gaussian profile in the  $z$ direction and are translationally invariant in the two directions  ${\bm x_\perp} = \{x_1,x_2\}$ orthogonal to $z$. Their
energy density per unit transverse area is 
$\mu^3(N_c^2/2\pi^2)$, with $\mu$ an arbitrary scale with respect to which all 
dimensionful quantities in the conformal  theory 
that we are working in can be measured.
The width $\sigma$ of the Gaussian energy-density profile of each sheet
is chosen to be $\sigma = 1/(2 \mu)$.

\begin{figure}
\vspace{-0.25in}
\hspace{-0.2in}\includegraphics[scale=0.38]{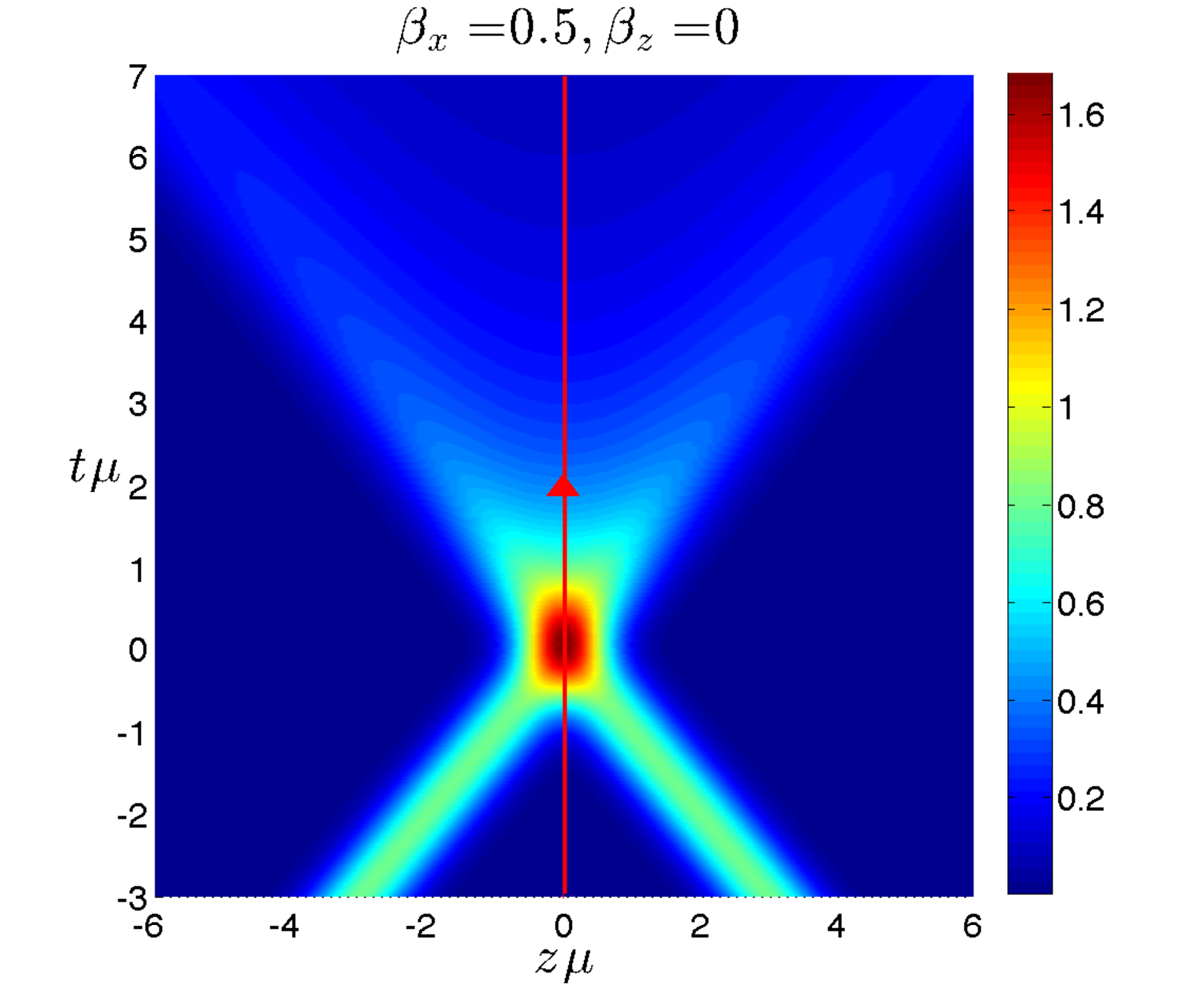} \hspace{-0.1in}
\includegraphics[scale=0.35]{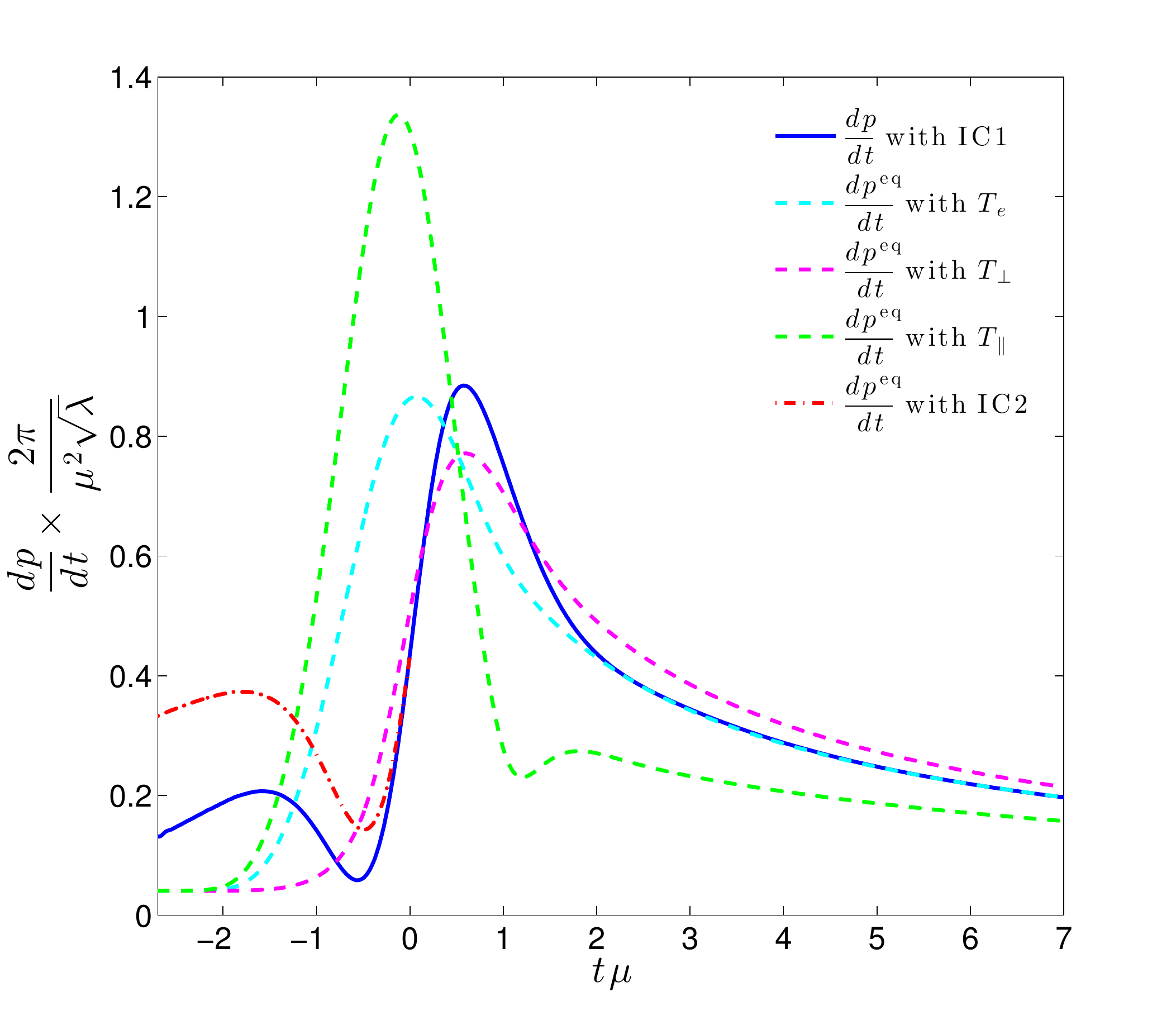}
\vspace{-0.25in}
\caption{Left: The path in $(t,z)$ of a quark moving in the $x$-direction with $\beta=0.5$ 
along $z=0$. Energy density of the colliding sheets (in units of $N_c^2 \mu^4/(2\pi^2)$) is shown by the color. Right: Instantaneous momentum loss of a heavy quark with two different
initial conditions for the string profile (solid blue and dot-dashed red curves). The results
agree for $t>0$, illustrating that our calculation of the momentum loss after the collision is insensitive
to the initial string profile.
The dashed cyan, magenta or green curves show what the momentum loss would be in an equilibrium
plasma with temperature $T_e$, $T_\perp$ or $T_\parallel$ determined from the instantaneous
energy density,  transverse pressure or parallel pressure.
}
\label{z0}
\end{figure}

\begin{figure}
\vspace{-0.25in}
\hspace{-0.25in}
\includegraphics[scale=0.37]{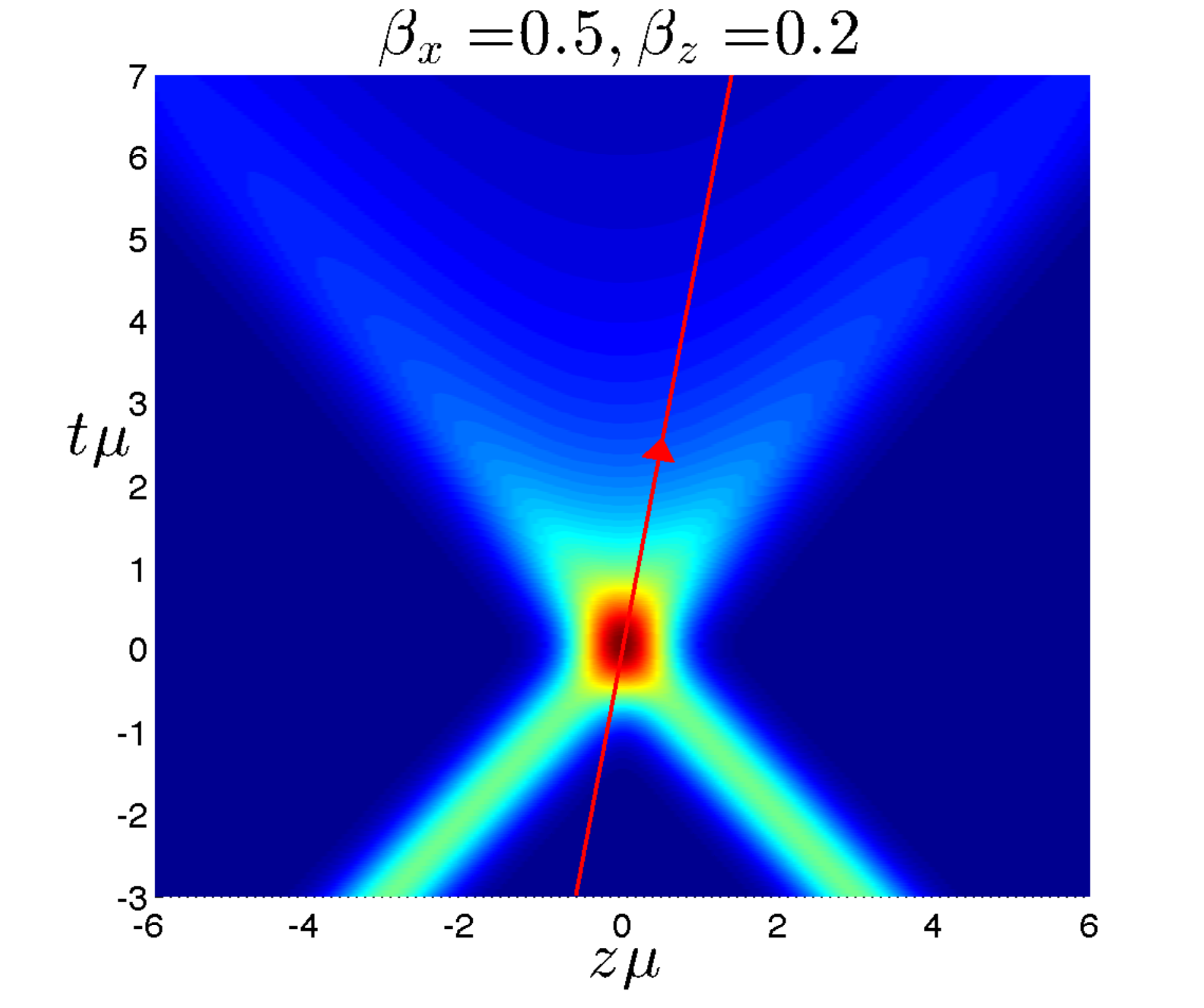}\hspace{-0.1in}
\includegraphics[scale=0.40]{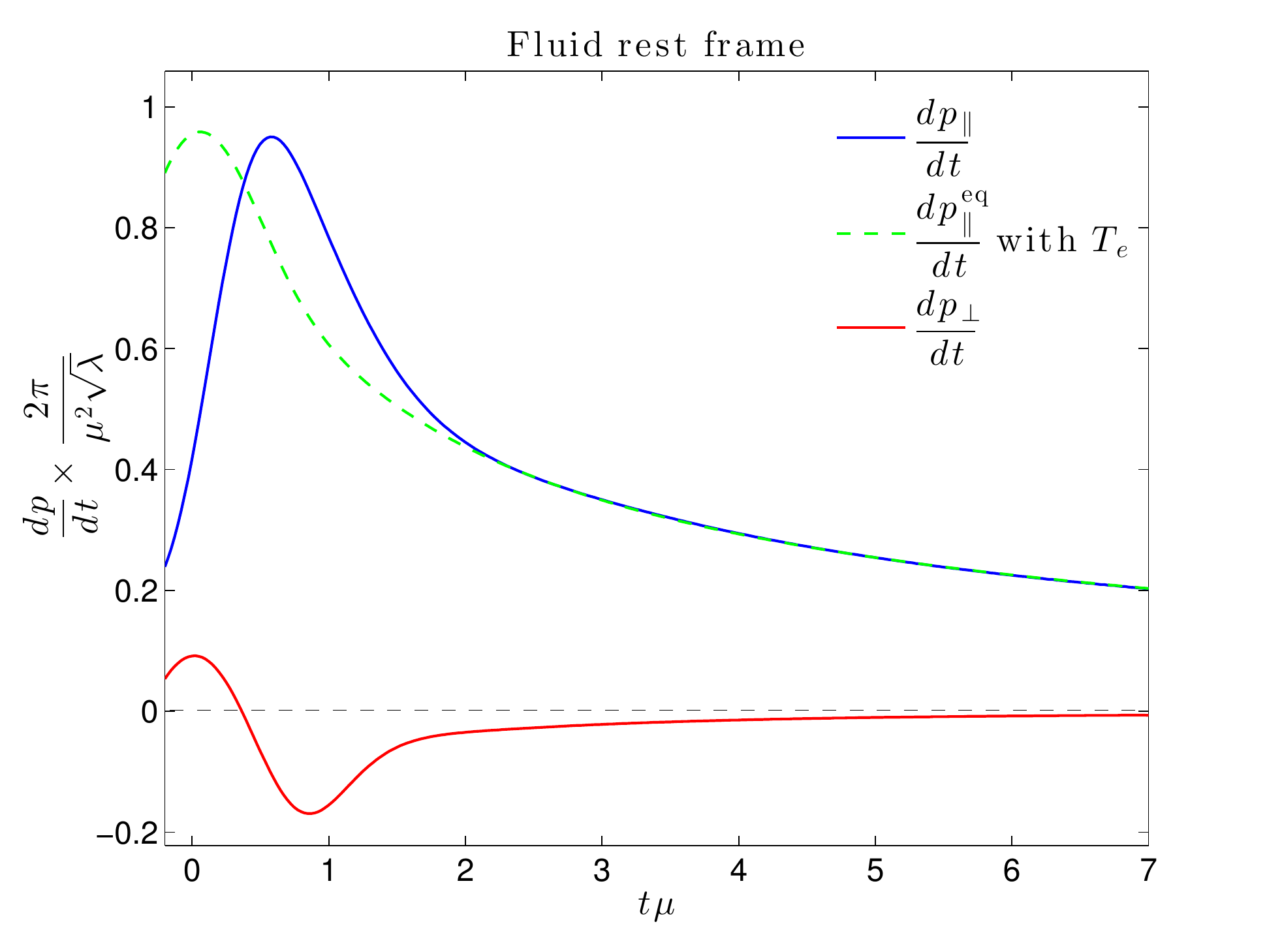}
\vspace{-0.25in}
\caption{Left: The path in $(t,z)$ of a quark with nonzero rapidity, specifically $\beta_z=0.2$
and $\beta_x=0.5$. Right: Instantaneous momentum loss in the fluid rest frame, antiparallel to the quark velocity (solid blue) and perpendicular to it (solid red). Green dashed curve shows the drag
force in an equilibrium plasma with temperature $T_e$.}
\label{znon0}
\end{figure}

We study heavy quark energy loss by
inserting a heavy quark moving at constant velocity $\vec \beta$ between the colliding sheets
before the collision and calculating 
the force needed to keep its velocity constant throughout the collision.
The energy density of the colliding sheets as well as two sample trajectories of a quark getting sandwiched between them are
shown in the left panels of Figs.~\ref{z0} and \ref{znon0}.
Via holography, the colliding planar sheets in SYM theory map into colliding planar gravitational 
waves in asymptotically AdS$_5$ spacetime~\cite{Chesler:2010bi}.   The addition of a heavy quark moving at constant velocity $\vec \beta$ amounts to including a classical 
string attached to the boundary of the geometry~\cite{Karch:2002sh} and dragging 
the string endpoint  at constant velocity $\vec \beta$, pulling the string
through the colliding gravitational wave geometry.  The force needed to maintain the velocity of the string endpoint yields the energy and momentum loss rates of the heavy quark~\cite{Herzog:2006gh}.


We employ infalling Eddington-Finkelstein coordinates to describe the asymptotically AdS$_5$ geometry.
Translation invariance in the $\bm x_{\perp}$ plane allows one to write the metric $G_{MN}$ 
in the form
\beq
ds^2 = G_{MN} dX^M dX^N = -A \,dv^2 + \Sigma^2\(e^B d \bm x_\perp^2 + e^{-2B} dz^2 \) 
+ 2\,F\, dv \,dz - 2\,du\, dv/u^2
\label{metric}
\eeq
where  $A, B, \Sigma$ and $F$ depend on coordinates $v$, $z$ and $u$.  Here $u$ is the AdS radial coordinate with the 
AdS boundary located at $u = 0$.  Lines of constant $v$ and $(z,{\bm x_{\perp}})$ are radially infalling null geodesics
affinely parameterized by $u$.  At $u=0$ the time coordinate $v$ coincides with the time coordinate $t$ in SYM. 
For initial conditions consisting of two colliding gravitational waves, 
the functions $A$, $B$, $F$ and $\Sigma$ were
obtained numerically in Ref.~\cite{Chesler:2010bi} by solving Einstein's equations.

We obtain the string equations of motion from the Polyakov action
\beq
S_{\rm P} = \int d\sigma d\tau \mathcal L_P = - \frac{T_0 L^2}{2} \int d\sigma d\tau \sqrt{-\eta} \eta^{ab} G_{MN} \partial_a X^M \partial_b X^N 
\label{Polyakov}
\eeq
where $T_0$ and $L$ are the string tension and AdS radius and where $\eta_{ab}$ is the worldsheet metric with $\tau$ and $\sigma$ being worldsheet coordinates
chosen such that 
\beq
\eta_{ab} = \left(
\begin{array}{cc}
  -\, \alpha(\tau,\sigma) & -1     \\
 -1 &      0
\end{array} \right)\,
\label{eta}
\eeq
with $\alpha(\tau,\sigma)$ to be determined.
With this choice of worldsheet coordinates,
which are the worldsheet analogue of infalling Eddington-Finkelstein coordinates,
lines of constant $\tau$ are infalling null worldsheet geodesics
and
the string equations of motion take the particularly simple form
\beq
\partial_\sigma \dot X^N + \Gamma_{AB}^N \dot X^A \partial_\sigma X^B = 0 \quad {\rm with} \quad \dot X^M \equiv \partial_\tau X^M 
- {\textstyle\frac{1}{2}}\,\alpha(\tau,\sigma)\, \partial_\sigma X^M\ ,
\label{eoms}
\eeq
where $\Gamma^N_{AB}$ are  Christoffel symbols.  
$\dot X^M$ is  the directional derivative along outgoing null worldsheet geodesics.  
The constraint equations 
are $(\partial_\sigma X)^2 = 0$ and $\dot X^2= 0$.  
The first is a temporal constraint: if it is satisfied at one value of $\tau$ and the other equations are satisfied, it remains satisfied at all times.  $\dot X^2= 0$ is a boundary constraint: if it is satisfied at one value of $\sigma$ and all other equations are satisfied, 
it is satisfied at all values of $\sigma$.  Residual worldsheet diffeomorphism invariance 
allows us to choose $u =  \sigma$.   Eqs.~\r{eoms} now prescribe the $\tau$-evolution of $X^M$, as follows: 
given $X^M$ at one $\tau$, use \r{eoms} to compute $\dot X^M$;
with $u=\sigma$ the definition of $\dot X^5$ becomes $\alpha= - 2\, \dot X^5$;
with $\alpha$ in hand, construct $\partial_\tau X^M$; finally, determine $X^M$ at the next
$\tau$.
To execute this procedure and obtain
$v(\tau,\sigma)$, $z(\tau,\sigma)$ and ${\bm x_{\perp}}(\tau,\sigma)$ 
we first need an initial profile of the string 
that satisfies $(\partial_\sigma X)^2 = 0$ and we need to
impose boundary conditions at $u=0$,
requiring that the endpoint of the string at $u=0$ moves at constant velocity $\vec \beta$
and that $\dot X^2= 0$ remains satisfied at $u=0$.  After we have solved \r{eoms}
and discovered how the string is buffeted by the colliding sheets,
we compute the canonical momentum flux $\pi^\sigma_\mu$ down the string at $u=0$.
This flux is minus the force required to maintain the velocity of the string endpoint 
and yields the four momentum lost by the quark per unit time 
\beq
\frac{dp_\mu}{dt} = \pi_\mu^\sigma  =  \frac{\sqrt{\lambda}}{2 \pi L^2} G_{\mu\nu} \bigl( \partial_\tau X^\nu - {\textstyle \alpha(\tau,\sigma) }\, \partial_\sigma X^\nu \bigr) \big |_{u=0}\ ,
\eeq
where $\lambda=g^2 N_c$ is the 't Hooft coupling in the gauge theory.
We choose several sets of initial conditions for the string and demonstrate that they all yield the same $dp_\mu/dt$ at suitably late times.


The instantaneous momentum loss $d\vec p/dt$ experienced
by the heavy quark  is given by the solid curves in Figs.~\ref{z0} and~\ref{znon0}. 
These curves are the central results of our calculation.
Fig.~\ref{z0} illustrates that 
we are initializing the string profile early enough that our results for the drag force are sensitive to how
we do this only  before the collision happens, not 
after the sheets collide ($t>0$). 
In Fig.~\ref{znon0}, 
the quark is 
at $z=0$ at $t=0$ but moves off the $z=0$ plane with some nonzero
rapidity.
In this case, in the frame in which we do the calculation the fluid 
is not at rest at the location of the quark for $t>0$.  However, in Fig.~\ref{znon0} we 
plot $d\vec p/dt$ in the
local fluid rest frame, meaning that at each instant in time we
transform to a frame in which the fluid at the location of the quark is static,
but of course not in equilibrium.
In particular, when the
quark is moving with nonzero rapidity the gradient
of the fluid velocity at the location of the quark is nonvanishing.
The blue curve shows the component of $d\vec p/dt$ in the direction opposite
to the velocity vector of the quark in the local fluid rest frame.  This 
is
a drag force, just as in the case of motion with zero rapidity in Fig.~\ref{z0}.
The red curve in Fig.~\ref{znon0} shows the component of $d\vec p/dt$ perpendicular
to the quark velocity in the local fluid rest frame.  A perpendicular force like this can
arise because of the presence of  gradients in the fluid velocity.  (We define the 
sign of the perpendicular component such that $dp_\perp/dt>0$ 
when $\left | dp_z^{\rm eq}/dt \right |$, see below, is greater than $\left | d p_z / dt \right | $.)


We are interested in comparing our result to expectations based upon
the classic result for the drag force required to move a heavy quark
with constant velocity $\vec \beta$ through the {\it equilibrium} plasma
in strongly coupled SYM theory, namely~\cite{Herzog:2006gh}
\beq
\left. \frac{d\vec p}{dt} \right |_{\rm eq} = \frac{\pi}{2} \sqrt{\lambda} T^2 \frac{\vec \beta}{\sqrt{1-\beta^2}}
\label{dpdt_dr}
\eeq
where $T$ is the temperature of the equilibrium plasma.
Out of equilibrium,
the matter does not have a single temperature.  We can nevertheless use (\ref{dpdt_dr}) to frame 
expectations for $d\vec p/dt$ at any point in spacetime, as follows. We transform to the local fluid
rest frame and define
 three  ``temperatures'' from the stress-energy tensor by writing it 
as $T_{\mu\nu}=(\pi^2 N_c^2/8)\,{\rm diag}\left(T_e^4,T_\perp^4,T_\perp^4,T_\parallel^4\right)$.
(In static equilibrium, $T_e=T_\perp=T_\parallel=T$.)  
We can then use
each of these three ``temperatures'' in the equilibrium expression 
(\ref{dpdt_dr}) and see to what degree the three curves that
result bracket our result for the energy loss of a heavy quark in far-from-equilibrium matter.  
We do this in Fig.~\ref{z0}.
The green dashed curve in Fig.~\ref{znon0} shows the force \r{dpdt_dr} in
an equilibrated plasma with $T=T_e$.
In an equilibrated plasma, the quark only feels
a drag force, opposing
its velocity;
the component of $d\vec p/dt$ perpendicular to the velocity of
the quark vanishes.  The red curve in Fig.~\ref{znon0}
arises because the heavy quark is propagating through out-of-equilibrium matter  
featuring velocity gradients.


With the exception of the (small) component of $d\vec p/dt$ 
depicted by the red curve in Fig.~\ref{znon0},
our results are rather well described
by the equilibrium expression (\ref{dpdt_dr}) with $T=T_e$.  At late times, we 
see from Figs.~\ref{z0} and $\ref{znon0}$ that the drag force that we have calculated
agrees quantitatively with this equilibrium expectation.  
This agreement begins at the same time that
viscous hydrodynamics starts to describe the
still far-from-isotropic fluid well~\cite{Chesler:2010bi}.
At early times, when the fluid is far from equilibrium, (\ref{dpdt_dr}) with $T=T_e$ does a reasonable
job of characterizing our results, although perhaps using $T=T_\perp$ does even better.  Certainly
there is no sign of any significant  ``extra'' energy loss arising by virtue of being far from equilibrium. 
The message of our calculation seems to be  that if we want to 
use (\ref{dpdt_dr}) to learn about heavy quark energy loss in heavy ion
collisions, it is reasonable to apply it throughout the collision, even before
equilibration, defining the $T$ that appears in it through the energy density.
The error that one would make by treating the far-from equilibrium energy loss
in this way is likely to be much smaller than other uncertainties.


\end{document}